\documentclass[twoside,english,american,aps,reprint,prl]{revtex4-1}
\usepackage[T1]{fontenc}
\usepackage{color}
\usepackage{babel}
\usepackage{units}
\usepackage{amstext}
\usepackage[unicode=true,pdfusetitle,
 bookmarks=true,bookmarksnumbered=false,bookmarksopen=false,
 breaklinks=true,pdfborder={0 0 0},backref=false,colorlinks=true]
 {hyperref}

\makeatletter
 \@ifundefined{textcolor}{}
 {%
   \definecolor{BLACK}{gray}{0}
   \definecolor{WHITE}{gray}{1}
   \definecolor{RED}{rgb}{1,0,0}
   \definecolor{GREEN}{rgb}{0,1,0}
   \definecolor{BLUE}{rgb}{0,0,1}
   \definecolor{CYAN}{cmyk}{1,0,0,0}
   \definecolor{MAGENTA}{cmyk}{0,1,0,0}
   \definecolor{YELLOW}{cmyk}{0,0,1,0}
 }

\usepackage[applemac]{inputenc}
 
\hypersetup{colorlinks=true,citecolor=blue,urlcolor=blue}

\makeatother

\begin{document}

\title{Multipartite monogamy of the concurrence}

\author{Marcio F. Cornelio}

\affiliation{Instituto de F\'isica, Universidade Federal de Mato Grosso, Cuiab\'a
- MT, 78068-900, Brazil}
\begin{abstract}
Monogamy of entanglement is generally discussed using a bipartite
entanglement measure as an upper bound. Here we discuss a new kind
of monogamous relation where the upper bound is given by a multipartite
measure of entanglement, the generalized concurrence. We show a new
monogamous equality involving the multipartite concurrence, all the
bipartite concurrences and the genuine tripartite entanglement for
pure three qubits system. The result extends to mixed states in an
inequality involving the generalized concurrence and all the bipartite
concurrences. We provide a counter-example showing that the result
cannot be extended for systems with more than three qubits. 
\end{abstract}
\maketitle
\emph{ }One of the most fundamental properties of quantum correlations
is that, differing from classical correlations, they are not shareable
at will when distributed among many parties. A simple example is a
pure maximally entangled state shared between Alice and Bob. This
state cannot share any additional correlation (classical or quantum)
with other parties. The composite system with a third party, say Carol,
can only be a tensor product of the state of Alice and Bob with the
state of Carol. This property has been called monogamy of entanglement
and there is a vast literature about \cite{Coffman2000,Christandl2004,Koashi2004,Osborne2006,Adesso2007,Cornelio2010a}.
The concept of monogamy is important for many tasks in the quantum
information theory, particularly, in the quantum cryptography \cite{Pawowski2010}.
In addition to the monogamy of entanglement, the concept of monogamy
has also appeared when discussing the violation of the Bell's inequalities
\cite{Pawowski2009a}. Also, it has been recently discussed in the
context of quantum correlations \cite{Streltsov2012} like the quantum
discord \cite{Ollivier2001}. Although, it has been shown that quantum
correlation measures cannot satisfy the traditional monogamous equation
\cite{Streltsov2012}, it has been shown it does satisfy a constrain
with entanglement of formation \cite{Bennett1996b,Cornelio2010b}
called conservation law working in a similar spirit to monogamy \cite{Fanchini2011}.
In this work, we will be restricted to the discussion of the monogamy
of entanglement. However, we shall see that, even in this case, one
can find new relations that go beyond the traditional monogamous inequality
(defined below in Eq. (\ref{eq:Monogamy3qubitsMIxed})). We find a
new monogamous equality involving the multipartite generalized concurrence
\cite{Mintert2005,Mintert2005a} which extends to an inequality for
mixed states. It seems to be the first time that monogamy is discussed
using a multipartite measure of entanglement as the upper bound to
other measures. In addition, we show that the result cannot be extend
to systems with more than three qubits. We believe that the reason
for this problem is that the generalized concurrence is not an additive
measure.

\emph{} The firsts to quantify this intuitive property of entanglement
were Coffman, Kundu and Wootters \cite{Coffman2000} in 2000. They
have shown that, for a pure three qubit state $\varphi_{abc}$, the
measure of entanglement called concurrence \cite{Bennett1996b,Wooters1998}
satisfies the following equality
\begin{equation}
C_{a|bc}^{2}(\varphi_{abc})=C^{2}(\rho_{ab})+C^{2}(\rho_{ac})+\tau,\label{eq:CKW-monogamy}
\end{equation}
where $C_{a|bc}$ stands for the concurrence between the subsystems
$S_{a}$ and $S_{bc}$ and $\tau$ stands for the genuine tripartite
entanglement. As the state $\varphi_{abc}$ is pure, the squared concurrence
$C_{a|bc}^{2}$ is just the linear entropy of the subsystem $S_{a}$,
\[
L(\rho_{a})=2(1-\textrm{Tr}\rho_{a}^{2}).
\]
The Eq. (\ref{eq:CKW-monogamy}) is understood in the following way:
the right side is the entanglement between the systems $S_{a}$ and
the composite system $S_{bc}$; this entanglement must be equal to
the entanglement between $S_{a}$ and $S_{b}$, plus the entanglement
between $S_{a}$ and $S_{c}$ and plus a genuine tripartite component
$\tau$. In this way, the entanglement between $S_{a}$ and $S_{bc}$
divides itself in three well understood parts. The genuine tripartite
entanglement must be interpreted as the entanglement that cannot be
captured by the bipartite concurrences and that goes away when any
one of the particles is lost. For instance, the state 
\[
\left|GHZ\right\rangle =\frac{1}{\sqrt{2}}\left(\left|000\right\rangle +\left|111\right\rangle \right)
\]
is the typical state with only genuine tripartite entanglement. For
this state, all the bipartite concurrences vanish and $\tau=1$. Also,
if any one of the parties is lost, the remaining two are completely
disentangled. On the other hand, the state
\[
\left|W\right\rangle =\frac{1}{\sqrt{3}}\left(\left|001\right\rangle +\left|010\right\rangle +\left|100\right\rangle \right)
\]
has only bipartite entanglement. In this case, $\tau=0$ and when
one of its parties is removed, the others two remain highly entangled.
In principle, there could be three distinct $\tau$'s depending on
which subsystem is chosen in the place of subsystem $S_{a}$ in Eq.
(\ref{eq:CKW-monogamy}). However, they turn out all equal for pure
three qubits systems \cite{Coffman2000}, showing a beautiful symmetry. 

However, this symmetry ends soon as we go for more general cases.
For mixed states, $\tau$ is not clearly defined and admits more than
one generalization \cite{Eltschka2008,Jung2009,Cao2010}. Indeed it
looses the symmetry with respect to the partition chosen to write
Eq. (\ref{eq:CKW-monogamy}). In addition, also the concurrence $C_{a|bc}^{2}$
has to be cleared, since it is uniquely defined only for pure states.
For the generalization of the concurrence, we take the usual convex
roof extension \cite{Coffman2000,Bennett1996b,Uhlmann2000} for the
linear entropy,
\begin{equation}
C_{a|bc}^{2}=\min_{{\cal E}}\left\{ \sum_{i}p_{i}L(\rho_{i}^{(a)})\right\} \label{eq:ConcurrenceHigherDimension}
\end{equation}
where the minimization runs over all the ensembles ${\cal E}$ of
pure states $\left\{ p_{i},\Phi_{i}\right\} $ such that $\rho_{abc}=\sum_{i}p_{i}\Phi_{i}$
and $\rho_{i}^{(a)}$ is the reduced state of the subsystem $S_{a}$
when the whole system, $S_{abc}$, is in the state $\Phi_{i}$. We
notice, however, that there are other possible generalizations of
the concurrence for higher dimensional systems \cite{Uhlmann2000,Rungta2001}.
With the definition (\ref{eq:ConcurrenceHigherDimension}) for the
concurrence, the following inequality is always true for any mixed
state $\rho_{abc}$ of three qubits systems \cite{Coffman2000},
\begin{equation}
C_{a|bc}^{2}(\rho_{abc})\ge C^{2}(\rho_{ab})+C^{2}(\rho_{ac}).\label{eq:Monogamy3qubitsMIxed}
\end{equation}
The inequality (\ref{eq:Monogamy3qubitsMIxed}) is considered the
fundamental monogamous inequality. Usually to say that a measure of
entanglement is monogamous means to say that it satisfies (\ref{eq:Monogamy3qubitsMIxed}).
When inequality (\ref{eq:Monogamy3qubitsMIxed}) is true for higher
dimensional systems, others inequalities can be derived. For example,
in Ref. \cite{Osborne2006}, it is shown that (\ref{eq:Monogamy3qubitsMIxed})
is true for the squared concurrence for systems of dimensions $2\times2\times N$
and, consequently, the following inequality for $N$ qubits systems
holds
\begin{equation}
C_{1|23\ldots N}^{2}\ge C_{12}^{2}+C_{13}^{2}+\cdots+C_{1N}^{2}.\label{eq:Monogamy}
\end{equation}
These inequalities can be called second order monogamy and other interesting
relations can be found \cite{Cornelio2010a}. However, most of the
measures of entanglement turns out not monogamous, that is, usually
not satisfying inequality (\ref{eq:Monogamy3qubitsMIxed}) in general,
with the exception of the squashed entanglement \cite{Christandl2004}.
The concurrence itself fails to satisfy (\ref{eq:Monogamy3qubitsMIxed})
for systems of dimensions greater or equal to $3\times3\times3$ \cite{Ou2007a}.
The entanglement of formation \cite{Bennett1996b} violates (\ref{eq:Monogamy3qubitsMIxed})
even for three qubit systems \cite{Coffman2000}. Nevertheless, other
measures of entanglement were proved monogamous in particular cases
\cite{Cornelio2010a,Koashi2004,Adesso2007}. \emph{ }

One of the main points of this paper is that the inequality (\ref{eq:Monogamy3qubitsMIxed})
does not exhaust the concept of monogamy.  In this work, we are going
to discuss a new kind of monogamy which we call multipartite. For
the best of our knowledge, all the monogamous relations put in the
right side of (\ref{eq:Monogamy3qubitsMIxed}) an upper bound defined
by a bipartite measure of entanglement. The main difference to our
new monogamous inequality is that the upper bound is defined by a
multipartite measure of entanglement and the left side involves the
bipartite entanglement between all the possible bi-partitions. We
will prove that such an inequality do exist for the concurrence in
three qubit systems. 

For that, first we need a generalization of the concurrence for multipartite
systems. The concurrence admits many generalizations for higher dimensional
and multipartite systems \cite{Uhlmann2000,Rungta2001,Mintert2005,Mintert2005a}.
Between them, the generalization most interesting for our purposes
here is the one by Mintert \emph{et al} \cite{Mintert2005,Mintert2005a}.
For a pure state $\Phi_{N}$ of an $N$ qubit system, it is defined
as
\begin{equation}
{\cal C}_{N}(\Phi_{N})=2^{1-\nicefrac{N}{2}}\sqrt{(2^{N}-2)-\sum_{i}\textrm{Tr}\rho_{i}}\label{eq:GeneralizedConcurence}
\end{equation}
where the summation runs over all the non-trivial subsystems of the
original $N$ qubit system. The interpretation of the generalized
concurrence is that it captures all kinds of entanglement in a given
system of $N$ particles. That is, it can be understood like the \textquotedbl{}sum\textquotedbl{}
of the bipartite entanglement over all bipartition, plus the tripartite
under all tripartition and so on. The main result of this work is
to quantify this interpretation for three qubit systems in the following
monogamous relation between the generalized concurrence, the bipartite
concurrences and the genuine tripartite entanglement\foreignlanguage{english}{,
\begin{equation}
{\cal C}_{3}^{2}(\Phi_{abc})=C^{2}(\rho_{ab})+C^{2}(\rho_{ac})+C^{2}(\rho_{bc})+\frac{3}{2}\tau.\label{eq:MMC}
\end{equation}
We can understand easily the Eq. (\ref{eq:MMC}). The right side is
a measure of all the entanglement contained in the state $\Phi_{abc}$
while the right side is the sum of the entanglement between all the
possible bi-partitions of the state plus the genuine tripartite entanglement
of the state. }

\selectlanguage{english}%
The proof of Eq. (\ref{eq:MMC}) is quite simple. First we rewrite
Eq. (\ref{eq:GeneralizedConcurence}) for $N=3$ as
\[
{\cal C}_{3}^{2}(\Phi_{abc})=\frac{1}{2}\left(6-\sum_{i}\textrm{Tr}\rho_{i}^{2}\right)
\]
which is
\[
{\cal C}_{3}^{2}(\Phi_{abc})=\frac{1}{2}\left[L(\rho_{a})+L(\rho_{b})+L(\rho_{c})\right],
\]
where $\rho_{i}$ is the reduced state of the subsystem $S_{i}$.
Using Eq. (\ref{eq:CKW-monogamy}) and its respective cyclic permutations
we get Eq. (\ref{eq:MMC}). One can check that Eq. (\ref{eq:MMC})
shares the same interpretation of Eq. (\ref{eq:CKW-monogamy}) with
respect to the states GHZ and W. For GHZ, all the concurrences vanish
and $\tau=1$, so 
\[
{\cal C}_{3}^{2}(\left|GHZ\right\rangle )=\frac{3}{2}.
\]
For the W state, we have that it is $\tau$ that vanishes and
\[
{\cal C}_{3}^{2}(\left|W\right\rangle )=C^{2}(\rho_{ab})+C^{2}(\rho_{ac})+C^{2}(\rho_{bc})=\frac{4}{3}.
\]

 We can generalize the result for mixed states. In this case, the
generalized concurrence is extended by the standard convex roof extension
\cite{Bennett1996b,Uhlmann2000,Mintert2005,Mintert2005a}. This process
is analogous to the one which is made in Eq. (\ref{eq:ConcurrenceHigherDimension}).
We define the squared generalized concurrence in Eq. (\ref{eq:MMC})
for a mixed state as
\begin{equation}
{\cal C}_{N}^{2}(\rho)=\min_{{\cal E}}\left\{ \sum_{i}p_{i}{\cal C}_{N}^{2}\left(\varphi_{i}\right)\right\} ,\label{eq:GCmixed}
\end{equation}
where the minimization goes over all the ensembles ${\cal E}$ of
pure states $\varphi_{i}$ such that $\rho=\sum_{i}p_{i}\varphi_{i}$
\footnote{This definition differs from the one originally given by Mintert \emph{et
al} \cite{Mintert2005} for the generalized concurrence of mixed states,
namely ${\cal C}_{N}(\rho)=\min_{{\cal E}}\left\{ \sum_{i}p_{i}{\cal C}_{N}\left(\varphi_{i}\right)\right\} .$
This difference resembles the one between the I-concurrence (without
the squared) and the tangle (with the squared) for the concurrence
of bipartite systems of dimensions higher than two. In principle,
none can be claimed better than the other, but the notion of the tangle
results in the CKW monogamy for mixed states and, here, in the multipartite
monogamy. %
}.  Now we take a ensemble that gives the minimum in Eq. (\ref{eq:GCmixed}).
For each $\varphi_{i}$ in this ensemble, the following inequality
follows from Eq. (\ref{eq:MMC}),
\[
{\cal C}_{3}^{2}(\varphi_{i})\geq C^{2}(\rho_{ab}^{i})+C^{2}(\rho_{ac}^{i})+C^{2}(\rho_{bc}^{i}).
\]
Making the summation over the ensemble, we have
\[
\sum_{i}p_{i}{\cal C}_{3}^{2}(\varphi_{i})\geq\sum_{i}p_{i}\left[C^{2}(\rho_{ab}^{i})+C^{2}(\rho_{ac}^{i})+C^{2}(\rho_{bc}^{i})\right].
\]
In the left side, we have ${\cal C}_{3}^{2}(\rho_{abc})$, since $\left\{ p_{i},\varphi_{i}\right\} $
is a minimizing ensemble. In the right side, from each $\varphi_{i}$
in the ensemble, we will get three density operators $\{\rho_{ab}^{i},\rho_{ac}^{i},\rho_{bc}^{i}\}$
which will occur with probability $p_{i}$. The right side is thus
the sum of three averages over the concurrences of those density matrices.
Each of these averages are made over an ensemble such that $\rho_{ab}=\sum_{i}p_{i}\rho_{ab}^{i}$
and so on. As the squared concurrence is a convex measure of entanglement,
we have 
\[
C^{2}(\rho_{ab})\le\sum_{i}p_{i}C^{2}(\rho_{ab}^{i})
\]
and so on for the other two terms. Therefore we have the following
inequality for every mixed state, $\rho_{abc}$, of a three qubit
system
\begin{equation}
{\cal C}_{3}^{2}(\rho_{abc})\ge C^{2}(\rho_{ab})+C^{2}(\rho_{ac})+C^{2}(\rho_{bc}).\label{eq:MonogamyMixed}
\end{equation}
The interpretation of the inequality (\ref{eq:MonogamyMixed}) is
clear. All the entanglement contained in the state $\rho_{abc}$ is
greater or equal to the sum of the entanglements between all possible
bi-partitions of the system. In addition, the inequality (\ref{eq:MonogamyMixed})
is also an easily calculable lower bound to the mixed state multipartite
concurrence. The generalized concurrence has no closed formula and
also no easy calculable lower bounds for multipartite systems. Therefore,
this lower bound (\ref{eq:MonogamyMixed}) is an additional very important
result \emph{per se} we have gotten from the multipartite monogamy
inequality (\ref{eq:MonogamyMixed}). 

One can ask whether inequality (\ref{eq:MonogamyMixed}) can be generalized
for systems of more than three qubits. The generalization of inequality
(\ref{eq:MonogamyMixed}) for an $N$ qubit system can be written
as
\begin{equation}
{\cal C}_{N}^{2}\geq\sum_{i<j}C_{ij}^{2}.\label{eq:MMCgeral}
\end{equation}
The simplest case is the one with four qubits. So we ask if the following
inequality is true
\begin{equation}
{\cal C}_{4}^{2}\geq C_{12}^{2}+C_{13}^{2}+C_{14}^{2}+C_{23}^{2}+C_{24}^{2}+C_{34}^{2}?\label{eq:MMC4qubits}
\end{equation}
Unfortunately the answer is no and the reason seems to be simple:
the generalized concurrence is not an additive measure. So, in order
of showing a counter-example, let us take four qubits in a state that
is a product of two maximally entangled states. In this case, the
left side of inequality (\ref{eq:MMC4qubits}) results
\[
{\cal C}_{4}^{2}=\frac{1}{2^{2}}\left(2^{4}-2-\sum_{i}\textrm{Tr}\rho_{i}^{2}\right)=\frac{1}{2^{2}}\left(14-7\right)=\frac{7}{4},
\]
while the right results
\[
C_{12}^{2}+C_{24}^{2}=2,
\]
which evidently violates the inequality (\ref{eq:MMC4qubits}). Therefore,
inequalities (\ref{eq:MMCgeral}) and (\ref{eq:MMC4qubits}) are not
true in general. This kind of counter-example is a rare event, however.
Using a computer, we have generated millions of typical states according
to the Haar measure and have not found any state violating the inequality
(\ref{eq:MMC4qubits}). This fact supports our guess that it is because
of the lack of the additivity property of the generalized concurrence
that the inequality (\ref{eq:MMC4qubits}) is false. Indeed, strong
connections between additivity and monogamy have already been found
in Ref. \cite{Cornelio2010a}. In addition, the only measure known
to be always monogamous is the squashed entanglement which, coincidently,
is also always additive \cite{Christandl2004,Koashi2004}.

In conclusion, we have shown a new kind of monogamy differing from
the traditional ones by the use of a multipartite measure of entanglement
as an upper bound to other measures. The discussion results in a beautiful
relation for the generalized concurrence for pure three qubit states.
The relation is extended to mixed states resulting in a monogamous
inequality that we have called multipartite monogamy. Unfortunately,
the result cannot be extended to more than three qubits. Nevertheless,
the concept of multipartite monogamy has proved its importance in
the validity of Eq. (\ref{eq:MMC}) and inequality (\ref{eq:MonogamyMixed}).
The fact that inequality (\ref{eq:MMCgeral}), the generalization,
is not always satisfied by the squared concurrence does not change
the fact that multipartite monogamy is a valid and interesting concept
for the study of multipartite entanglement. We expect that this work
may open a new branch in the research for monogamous relations involving
multipartite measures of entanglement. The study of multipartite measures
is still in its beginnings and more measures are probably to come.
The concept of monogamy here developed will probably be important
for the judgment of good measures of multipartite entanglement. The
Eq. (\ref{eq:MMC}) and inequality (\ref{eq:MonogamyMixed}) may be
the firsts of a series of new monogamous relations that multipartite
measures should satisfy. In this sense, we believe that this work
goes beyond showing a new monogamous relations, it opens a new branch
of research for monogamy and multipartite entanglement. 

%\emph{Acknowledgements} - This work is supported by CNPq through the
%National Institute for Science and Technology of Quantum Information
%(INCT-IQ).

\bibliographystyle{apsrev4-1}
\bibliography{/Users/marcio/Documents/bibtex/MMC,Additividademulti}
\selectlanguage{american}%

\end{document}